\documentclass[aps,pre,english,showpacs,preprintnumbers,twocolumn,floatfix,
nofootinbib,10pt]{revtex4-1}
\usepackage{eurosym}
\usepackage{amssymb}
\usepackage{graphicx,color}
\usepackage{epsfig}
\usepackage{rotating}
\usepackage[T1]{fontenc}
\usepackage{latexsym,epsfig}
\usepackage[latin9]{inputenc}
\usepackage{amsfonts}
\usepackage{dcolumn}
\usepackage{bm}
\usepackage{babel}
\usepackage{hyperref}
\usepackage{amsmath,mathrsfs}

\begin{document}

\title{An alternative method to characterize first- and second-order phase
transitions in surface reaction models.}
\author{Henrique A. Fernandes$^1$, Roberto da Silva$^2$, Eder D. Santos$^1$%
,Paulo F. Gomes$^1$, Everaldo Arashiro$^3$}

\affiliation{$^1$Coordena{\c{c}}{\~a}o de F{\'i}sica, Universidade Federal de Goi{\'a}s, Regional Jata{\'i}, BR 364, km 192, 3800 - CEP 75801-615, Jata{\'i}, Goi{\'a}s, Brazil\\
$^2$Instituto de F{\'i}sica, Universidade Federal do Rio Grande do Sul, Av. Bento Gon{\c{c}}alves, 9500 - CEP 91501-970, Porto Alegre, Rio Grande do Sul, Brazil\\
$^3$Instituto de Matem{\'a}tica, Estat{\'i}stica e F{\'i}sica, Universidade Federal do Rio Grande, Campus Carreiros, Av. It{\'a}lia, km 8 - CEP 96203-900, Rio Grande, Rio Grande do Sul, Brazil}

\keywords{ZGB model, Refinement process, Models with absorbent steady
state,Time-dependent Monte Carlo simulation}

\pacs{05.10.-a; 02.70.Tt, 05.70.Ln} 

\begin{abstract}
In this work, we revisited the Ziff-Gulari-Barshad (ZGB) model to study its
phase transitions and critical exponents through time-dependent Monte Carlo
simulations. We used a method proposed recently to locate the
non-equilibrium second-order phase transitions and that has been
successfully used in systems with defined Hamiltonians and with absorbing
states. This method, which is based on optimization of the coefficient of
determination of the order parameter, was able to characterize the
second-order phase transition of the model, as well as its upper spinodal
point, a pseudo-critical point located near the first-order transition. The
static critical exponents $\beta $, $\nu_{\parallel }$, and $\nu_{\perp }$,
as well as the dynamic critical exponents $\theta $ and $z$ for the
second-order point were also estimated and are in excellent agreement with
results found in literature.
\end{abstract}

\maketitle

\section{Introduction}

\label{introduction}

In recent years, the study of kinetic or nonequilibrium systems \cite%
{OliveiraTome,hinrichsen2000} has grown considerably making them a fruitful
subject in some branches of the biological \cite{Zia2011,AlvesJr2005},
financial \cite{Ingber1984,Zembrzusky2010}, social \cite%
{Castellano2009,socialrob} and applied sciences \cite{Luzzi2010,aplliedrob}.
Major efforts and interest have been put on systems which exhibit
nonequilibrium phase transitions and critical phenomena such as transport
phenomena, traffic jams, and epidemic spreading \cite{hinrichsen2000a}. In
this context, one can also consider the directed percolation (DP) \cite%
{blease1977}: an important case of nonequilibrium critical phenomena whose
class cover other interesting models with universal exponents. Systems
belonging to DP universality class exhibit a second-order phase transition
from an active phase to an absorbing phase. The absorbing phase represents
states in which, once reached, the systems become trapped and can not
escape. As conjectured by Janssen \cite{janssen1981} and Grassberger \cite%
{grassberger1982} there exist many physical systems belonging to the DP
universality class. Nevertheless, experimental observations of such a
behavior have not been shown frequently in the literature.

Other nonequilibrium systems which present phase transitions and critical
phenomena are related to surface reaction models \cite%
{evans1991a,albano1994,andrade2010,andrade2012}. In fact, these models have
attracted considerable interest whereas they can be used to explain several
experimental observations in catalysis \cite%
{ehsasi1989,christmann1991,imbhil1995}. For instance, in 1986, Ziff, Gulari
and Barshad \cite{ziff1986} devised a stochastic model that describes some
nonequilibrium aspects of the catalytic reaction of carbon monoxide and
oxygen to produce carbon dioxide ($CO + O \rightarrow CO_2$) on a surface
and that, in addition, exhibits second- and first-order phase transitions.
Several works have shown that its critical point belongs to the DP
universality class \cite{jensen1990}. Due to its simplicity, rich phase
diagram, and experimental observation of the first-order phase transition,
the Ziff-Gulari-Barshad model, also know as ZGB model, has become a
prototype for the study of reaction processes on catalytic surfaces \cite%
{meakin1987,dickman1986, fischer1989}.

After its advent, a number of authors have proposed some modified versions
of the ZGB model in order to obtain more realistic systems of actual
catalytic processes. For instance, it was modified to include $CO$
desorption \cite{fischer1989,dumont1990,jensen1990a,albano1992,tome1993,
kaukonen1989}, diffusion \cite{ehsasi1989,fischer1989,kaukonen1989,
jensen1990a,grandi2002}, impurities \cite{hoenicke2000,buendia2012,
buendia2013,hoenicke2014,buendia2015}, attractive and repulsive interactions
between the adsorbed molecules \cite{satulovsky1992}, surfaces of different
geometries \cite{meakin1987,albano1990} and with hard oxygen boundary
conditions \cite{brosilow1993}, etc. In addition, it has been studied
through several techniques, such as simulations, mean-field theories, series
analysis, etc \cite{marro1999}.

In this manuscript, we revisit the ZGB model as proposed by Ziff, Gulari,
and Barshad in 1986 \cite{ziff1986}, in order to study its phase transitions
and critical exponents by using short-time dynamics. By considering
time-dependent Monte Carlo simulations and a non-conventional optimization
method, based on a simple statistical concept known as coefficient of
determination (see, for example, Ref. \cite{trivedi2002}), we were able to
refine the second-order transition point and, surprisingly, obtain an
accurate estimate to the upper spinodal point associated to the first-order
transition point. This technique has been used in the study of reversible
systems \cite{roberto2012, roberto2013a,roberto2013b,roberto2014} and was
considered recently in the study of an epidemic model to determine its
critical immunization probability \cite{roberto2015}.

The paper is organized as follows: In the next section, we present the model
and in Section \ref{sec:method} we describe the short-time Monte Carlo
simulation technique as well as the coefficient of determination. In Section %
\ref{sec:results}, we show our main numerical results and illustrations.
Finally, a brief summary is presented in Section \ref{conclusions}.

\section{Ziff-Gulari-Barshad model}

\label{sec:model}

The Ziff-Gulari-Barshad (ZGB) model \cite{ziff1986} is a dimer-monomer
lattice model which simulates the catalysis between the carbon monoxide $%
(CO) $ and the oxygen molecule $(O_2)$. The reactions follow the
Langmuir-Hinshelwood mechanism \cite{ziff1986,evans1991b} and are summarized
in three steps, as follows: 
\begin{eqnarray}
CO(g)+V \rightarrow CO(a),  \label{eq01} \\
O_2(g)+2V \rightarrow 2O(a),  \label{eq02} \\
CO(a)+O(a) \rightarrow CO_2(g)+2V,  \label{eq03}
\end{eqnarray}
where $g$ and $a$ refer, respectively, to the gas and adsorbed phases of the
atoms/molecules, $CO_2$ stands for carbon dioxide molecule, and $V$ means a
vacant site on the surface.

Computationally, this catalytic surface can be modeled as a regular square
lattice and its sites might be occupied by $CO$ molecules or by oxygen ($O$)
atoms or may be empty. By using the Monte Carlo method, the simulation is
carried out as follows \cite{albano1996,marro1999,ziff1986}: In the gas
phase, the $CO$ molecule is chosen to impinge on the surface at rate $y_{CO}$%
, while the $O_{2}$ molecule strikes the lattice at rate $y_{O_{2}}=1-y_{CO}$%
. As these rates are relative ones and $y_{CO}+y_{O_{2}}=1$, the model has a
single free parameter: $y=y_{CO}$. According to Eq. (\ref{eq01}), if the $CO$
molecule is selected in the gas phase, a site on the surface is chosen at
random and, if it is vacant ($V$), the molecule is adsorbed on this site.
Otherwise, if the chosen site is occupied by a $CO$ molecule or by an $O$
atom, the trial ends, the $CO$ molecule returns to the gas phase, and a new
molecule is chosen. However, if the $O_{2}$ molecule is selected, a
nearest-neighbor pair of sites is chosen at random. If both sites are
vacant, the $O_{2}$ molecule dissociates into a pair of $O$ atoms and are
adsorbed on the chosen lattice sites [Eq. (\ref{eq02})]. Otherwise, if one
or both sites are occupied, the trial ends, the $O_{2}$ molecule returns to
the gas phase, and a new molecule is chosen. Eq. (\ref{eq03}) stands for the
reaction between the $CO$ molecule and the $O$ atom, both adsorbed in the
lattice. Immediately after each adsorption event, the nearest-neighbor sites
of the adsorbed molecule are checked. If a $O-CO$ pair is found, the $CO_{2}$
molecule is formed and quits the lattice, leaving two vacant sites on it.
However, if there is the formation of two or more $O-CO$ pairs, a pair is
chosen at random to quit the lattice.

The ZGB model has been vastly studied and nowadays is considered a prototype
for the study of reaction processes on catalytic surfaces. This is mainly
due to its simplicity and rich phase diagram with three distinct
steady-state phases separated by second- and first-order phase transitions 
\cite{ziff1986,meakin1987,dickman1986,fischer1989}. For $0 < y < y_1$ the
surface becomes irreversibly poisoned (saturated) by $O$ atoms ($O-$poisoned
state). At $y = y_1 \cong 0.3874$ \cite{voigt1997} there is a second-order
phase transition from the $O-$poisoned state to an active phase where there
is sustainable production of $CO_2$ molecules. This state ends when $y = y_2
\cong 0.5256$ \cite{ziff1992} whereas for this point the system undergoes a
first-order phase transition and the surface becomes irreversibly poisoned
by $CO$ molecules ($CO-$poisoned state). For $y_2 < y \leq 1$ the surface
remains in the $CO-$poisoned state, i.e., every site on the surface is
occupied by $CO$. In summary, $y_1$ and $y_2$ are irreversible phase
transition (IPT) points between the reactive and poisoned states. While $y_1$
is related to the second-order IPT, $y_2$ represents the first-order one.
Although some experimental works on platinum confirm the existence of
first-order transition in the catalytic oxidation of $CO$ \cite%
{golchet1978,matsushima1979,ehsasi1989,christmann1991,block1993, berdau1999}%
, there is no experimental evidences of second-order IPT despite its
existence in the theoretical framework. In this case, it is well established
that this transition belongs to the DP universality class \cite%
{grinstein1989,jensen1990}.

\section{Finite size scaling and time-dependent Monte Carlo simulations}

\label{sec:method}

The finite size scaling near criticality of systems belonging to the DP
universality class can be described by: 
\begin{equation}
\left\langle \rho(t)\right\rangle \sim
t^{-\beta/\nu_{\parallel}}f((y-y_{c})t^{1/\nu_{\parallel}},t^{d/z}L^{-d},%
\rho_{0}t^{\beta/\nu_{\parallel}+\theta}),  \label{eqfss}
\end{equation}
where $\left\langle \cdots \right\rangle$ means the average on different
evolutions of the system, $d$ is the dimension of the system ($d=2$ for the
ZGB model), $L$ is its linear size, and $t$ is the time. The exponents $%
z=\nu_{\parallel}/\nu_{\perp}$ and $\theta=\frac{d}{z}-\frac{2\beta}{%
\nu_{\parallel}}$ are dynamic critical exponents, and $\beta$, $%
\nu_{\parallel}$, and $\nu_{\perp}$ are static ones. Here, $y-y_{c}$ denotes
the distance of a point $y$ to the critical one point, $y_{c}$, which
governs the algebraic behaviors of the two independent correlation lengths:
the spatial one which behaves as $\xi_{\perp}\sim (y-y_c)^{-\nu_{\perp}}$
and temporal one, $\xi_{\parallel}\sim (y-y_c)^{-\nu_{\parallel}}$.
Basically, $\xi_{\perp}$ must be thought of as the average over many
independent realizations of the cluster diameter while $\xi_{\parallel}$ is
the same average of the required time to reach the absorbing state. Besides
the density of $CO$ molecules $\rho_{CO}$, one can also consider the density
of empty (vacant) sites $\rho_V $ as the order parameter of the model.
Therefore, in Eq. (\ref{eqfss}), $\rho$ stands for a generic density which
can be $\rho_{CO}$ or $\rho_V$. The density is given by 
\begin{equation*}
\rho(t)=\frac{1}{L^{d}}\sum_{j=1}^{L^{d}} s_{j}.
\end{equation*}
According to the density which is taken into consideration, $s_{j}=1$ when
the sites $j$ are occupied by $CO$ molecules (for $\rho_{CO}$) or when they
are vacant (for $\rho_{V}$). Otherwise, the $s_j=0$. As can be seen in Sec. %
\ref{sec:results}, part of our results are obtained by considering both
order parameters.

The dynamic and static critical exponents of the model can be obtained by
using the Eq. (\ref{eqfss}) and performing time-dependent Monte Carlo
simulations with two different initial conditions. Eq. (\ref{eqfss}) can be
observed in another way:

\begin{equation*}
\left\langle \rho \right\rangle (t,L,\rho _{0})=L^{-\beta /\nu _{\perp
}}\left\langle \rho \right\rangle (L^{-z}t,L^{x_{0}}\rho _{0})
\end{equation*}
where $x_{0}=\beta /\nu_{\perp }+z\theta $ at $y=y_{c}$. Denoting $u=tL^{-z} 
$ and $w=L^{x_{0}}\rho_{0}$, the derivative with respect to $L$ gives

\begin{eqnarray*}
\partial_{L}\left\langle \rho \right\rangle &=&(-\beta/\nu_{\perp})
L^{-\beta/\nu_{\perp}-1}\left\langle \rho \right\rangle (u,w) \\
&&+L^{-\beta/\nu_{\perp}}[\partial_{u}\left\langle \rho \right\rangle
\partial_{L}u+\partial_{w}\left\langle \rho \right\rangle \partial _{L}w],
\end{eqnarray*}
where one have explicitly $\partial_{L}u=-ztL^{-z-1}$ and $%
\partial_{L}w=x_{0}\rho_{0}L^{x_{0}-1}$. In the limit $L\rightarrow \infty$,
which implicates in $\partial_{L}\left\langle \rho \right\rangle \rightarrow
0$, one has $x_{0}w\partial _{w}\left\langle \rho
\right\rangle-zu\partial_{u}\left\langle \rho \right\rangle
-\beta/\nu_{\perp}\left\langle \rho \right\rangle =0$. The separability of
the variables $u$\ and $w$, i.e., $\left\langle \rho \right\rangle
(u,w)=\left\langle \rho \right\rangle_{u}(u)\left\langle \rho
\right\rangle_{w}(w)$\ leads to 
\begin{equation*}
x_{0}w\left\langle \rho \right\rangle_{w}^{\prime}/\left\langle \rho
\right\rangle_{w}=\beta/\nu_{\perp }+zu\left\langle \rho \right
\rangle_{u}^{\prime}/\left\langle \rho \right\rangle_{u}=c,
\end{equation*}
where $c$ must be equal to a constant. So, we have $\left\langle \rho
\right\rangle_{u}=u^{c/z}-\beta/(\nu _{\perp }z)$ and $\left\langle \rho
\right\rangle _{w}=w^{c/x_{0}}$, which leads to: 
\begin{equation*}
\left\langle \rho \right\rangle(t)= \rho_{0}^{c/x_{0}}t^{(c-\beta/\nu_{\perp
})/z}.
\end{equation*}
When one considers the system starting with all sites empty, there is no
dependence on initial conditions ($c=0$) and 
\begin{equation}
\left\langle \rho \right\rangle (t)\sim t^{-\beta/\nu_{\parallel}}.
\label{eq_pl1}
\end{equation}
However, when the simulation starts with all sites of the lattice filled
with $O$ atom but a random site which remains empty, we can choose $c=x_{0}$
which leads to 
\begin{equation}
\left\langle \rho \right\rangle(t)\sim \rho_{0}t^{\theta}= \rho_{0}t^{\left(%
\frac{d}{z}-2\frac{\beta}{\nu_{\parallel }}\right)}.  \label{eq_pl2}
\end{equation}
Here it is important to notice an interesting crossover phenomena \cite%
{hinrichsen2000}. By starting with an initial density $\rho_{0}$, the
density of active sites (empty sites in our case) increases as shown in Eq. (%
\ref{eq_pl2}). This phenomena is known as the critical initial slip of
non-equilibrium systems and occurs until it reaches a maximum value at time $%
t_{\max}$. Thereafter, the system cross over to the usual relaxation
described by the power law decay given by Eq. (\ref{eq_pl1}). In summary: 
\begin{equation*}
\left\langle \rho \right\rangle (t)=\left\{%
\begin{array}{lll}
\rho _{0}t^{\theta} & \text{if} & t<t_{\max } \\ 
&  &  \\ 
t^{-\beta /\nu _{\parallel }} & \text{for} & t>t_{\max }%
\end{array}%
\right.
\end{equation*}%
where $t_{\max}$ is the solution of $\rho_{0}t_{\max}^{\theta}=
t_{\max}^{-\beta/\nu_{\parallel}}$ which gives $t_{\max }=\rho_0^{-1/\left(%
\frac{d}{z}-\frac{\beta }{\nu _{\parallel }}\right)}$. Such a relaxation is
similar to that one which occurs for spin systems when they are quenched
from high temperature to the critical one \cite{janssen1989}.

An interesting way to obtain the exponent $z$ from an independent way is to
combine simulations with different initial conditions. This idea has been
applied successfully in a large number of spin systems: for example, the
Ising model, the $q=3$ and $q=4$ Potts models \cite{Silva20021}, Heisenberg
model \cite{Fernandes2006c} and even for models based on generalized Tsallis
statistics \cite{Silva2012-Tsallis}, was introduced recently in systems
without defined Hamiltonian, as can be seen in Ref. \cite{Roberto2004}. To
obtain the power law, we consider the cumulant as follows: 
\begin{equation}
F_{2}(t)=\frac{\left\langle \rho \right\rangle _{\rho _{0}=1/L}(t)}{%
\left\langle \rho \right\rangle _{\rho _{0}=1}^{2}(t)}\sim t^{d/z}.
\label{eq_f2}
\end{equation}%
So, once the dimension $d$ of the system is known, a $\text{log}-\text{log}$
fit of $F_{2}(t)\times t$ yields the exponent $z$.

In addition to the exponents $z$ and $\theta $ which are obtained
independently from Eqs. (\ref{eq_pl2}) and (\ref{eq_f2}), we can obtain the
static critical exponents $\beta $, $\nu _{\parallel }$, and $\nu _{\perp }$
by using the method proposed by Grassberger and Zhang \cite{Grassberger1996}
to estimate the exponent $\nu _{\parallel }$ for DP and used by da Silva 
\textit{et al.} \cite{Roberto2004} to study the one-dimensional contact
process and Domany-Kinzel cellular automaton through short-time Monte Carlo
simulations, 
\begin{equation}
D(t)=\frac{\partial \ln \left\langle \rho \right\rangle }{\partial y}%
\bigg\vert_{y=y_{c}}=t^{\frac{1}{\nu _{\parallel }}}.  \label{eq_dt}
\end{equation}%
Here, the derivative is numerically represented by 
\begin{equation*}
D(t)=\frac{1}{2\delta }\text{ln}\left( \frac{\left\langle \rho \right\rangle
(y_{c}+\delta )}{\left\langle \rho \right\rangle (y_{c}-\delta )}\right) ,
\end{equation*}%
where $\delta $ is a tiny perturbation needed to move the system slightly
off the criticality.

\section{Results}

\label{sec:results}

Nonequilibrium Monte Carlo simulations was first designed to study
second-order critical points whereas, at these points, universality and
scaling behavior is observed even at the early stages of time evolution \cite%
{janssen1989,huse1989}. However, it has been shown that this technique is
also important in the study of weak first-order phase transitions \cite%
{Schulke2001,Albano2001} since these transitions possess long correlation
lengths and small discontinuities and therefore behave similarly to
second-order phase transitions. It has been conjectured that near a weak
first-order transition there exist two pseudo-critical points: one point is
just below (inferior) the first-order point, and the other is just above
(upper) it. These pseudo-critical points are known as spinodal points.

In this contribution, we divide our results in two parts. First, we perform
nonequilibrium Monte Carlo simulations to characterize the first- and
second-order transitions of the model. For this task, we use an alternative
method based on optimization of the coefficient of determination of power
laws. Surprisingly, we obtain a description of the upper spinodal point
which has not been observed by this method, developed by one of authors in
2012 \cite{roberto2012}, for models without defined Hamiltonian.

In the second part of our results, we carry out short-time Monte Carlo
simulations to determine the static critical exponents $\beta,
\nu_{\parallel}$, $\nu_{\perp}$, and the dynamic critical exponents $z$ and $%
\theta$ of the second-order point of the ZGB model using a set of power
laws. In this study, we show that the method of mixed initial conditions
applied to other models without defined Hamiltonian \cite{Roberto2004}
(contact process, cellular automata) can be adapted also to obtain the
dynamic exponent $z$ of the ZGB model.

\subsection{Results I: Exploration of the upper spinodal point and
second-order transition point using the coefficient of determination}

\label{subsec:results1}

The main goal of this work is to study the phase transition points of the
ZGB model via time-dependent Monte Carlo (MC) simulations by estimating the
best $y$ given as input the parameter $y^{(\min)}$ (initial value) and run
simulations for different values of $y$ up to $y^{(\max)}$, according to a
resolution $\Delta y$.

For this task, we used an approach developed in Ref. \cite{roberto2012} in
the context of generalized statistics. This tool had also been applied
successfully to study multicritical points, for example, tricritical points 
\cite{roberto2013b} and Lifshitz point of the ANNNI model \cite{roberto2013a}%
, Z5 model \cite{roberto2014} and also in models without defined Hamiltonian 
\cite{roberto2015}.

Since at criticality ($y=y_{c}$) it is expected that the order parameter
obeys the power law behavior of Eq. (\ref{eq_pl1}), we performed MC
simulations for each value of $y=y^{(\min)}+i\Delta y$, with $i=1,...,n$,
where $n=\left\lfloor (y^{(\max)}-y^{(\min )})/\Delta y\right\rfloor$, and
calculated the coefficient of determination, which is given by 
\begin{equation}
r=\frac{\sum\limits_{t=1}^{N_{MC}}(\overline{\ln \left\langle \rho
\right\rangle }-a-b\ln t)^{2}}{\sum\limits_{t=1}^{N_{MC}}(\overline{\ln
\left\langle \rho \right\rangle }-\ln \left\langle \rho \right\rangle
(t))^{2}}\text{,}  \label{determination_coefficient}
\end{equation}
where $\overline{\ln \left\langle \rho \right\rangle} =(1/N_{MC})\sum%
\nolimits_{t=1}^{N_{MC}}\ln \left\langle \rho \right\rangle(t)$, and the
critical value $y_{c}$ corresponds to $y^{(opt)}=\arg \max_{y\in \lbrack
y^{(\min )},y^{(\max )}]}\{r\}$. The coefficient $r$ has a very simple
explanation: it measures the ratio: (expected variation)/(total variation).
The bigger the $r$, the better the linear fit in log-scale, and therefore,
the better the power law which corresponds to the critical parameter except
for an order of error $\Delta y$.

It is important to mention that the coefficient of determination was
obtained by considering two order parameters: the density of $CO$ molecules (%
$\rho_{CO}$) and the density of empty sites $\rho_{V}$. Although the former
is commonly used, some studies have considered $\rho_{V}$ as order parameter
(see Ref. \cite{Leite2001}). First, we considered a lattice of linear size $%
L=160$ and explored the scenery in general by estimating $r$ for different
values of $y$ ($0.3 \leq y \leq 0.6$ and $\Delta y=10^{-4}$) (see Fig. \ref%
{Fig:Optimization}). The square (red) points represent the coefficient of
determination obtained when considering $\rho_{V}$ while the circles (blue)
represent the coefficient of determination for $\rho_{CO}$.

\begin{figure}[th]
\begin{center}
\includegraphics[width=\columnwidth]{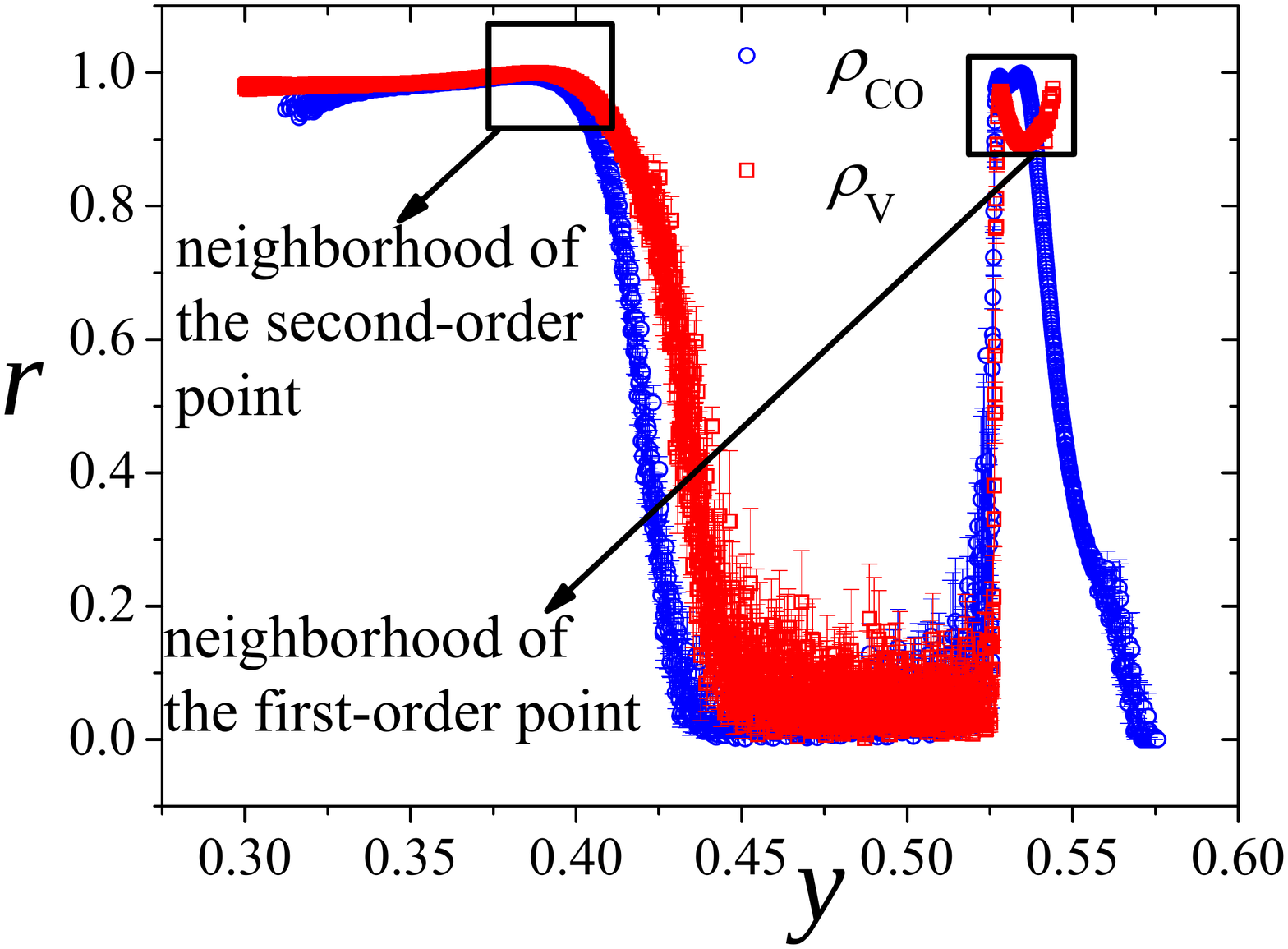}
\end{center}
\caption{Coefficient of determination $r$ as funcion of $y$. The maximum
occur at the expected critical point and in a region related to the
first-order transition which is also observed in our study. Both regions
deserve our attention and are explored in this paper.}
\label{Fig:Optimization}
\end{figure}
As can be seen in this figure, when one takes into account the density of $%
CO $ molecules the curve ends for $y\simeq 0.56$. However, when one
considers the density of empty sites, the curve ends for $y\simeq 0.55$. The
reason is that for higher values of $y$, one obtains undefined values
meaning that there is no power law behavior as observed in second-order
phase transitions and the slope goes to infinity. We can also observe two
candidate regions to have phase transitions ($r\simeq 1$): one maximum for
the expected critical point ($y\simeq 0.3874$) and a region related to the
first-order transition ($y\simeq 0.525$). Of course, both regions deserve
our attention. Hence, we explore such parts by performing simulations for
each region, with $\Delta y=10^{-4}$ and for different lattice sizes ($L=40$%
, 80, 160, 240, and 320). For each lattice, the process was repeated for
five different seeds in order to obtain the error bars.

\begin{figure}[th]
\begin{center}
\includegraphics[width=\columnwidth]{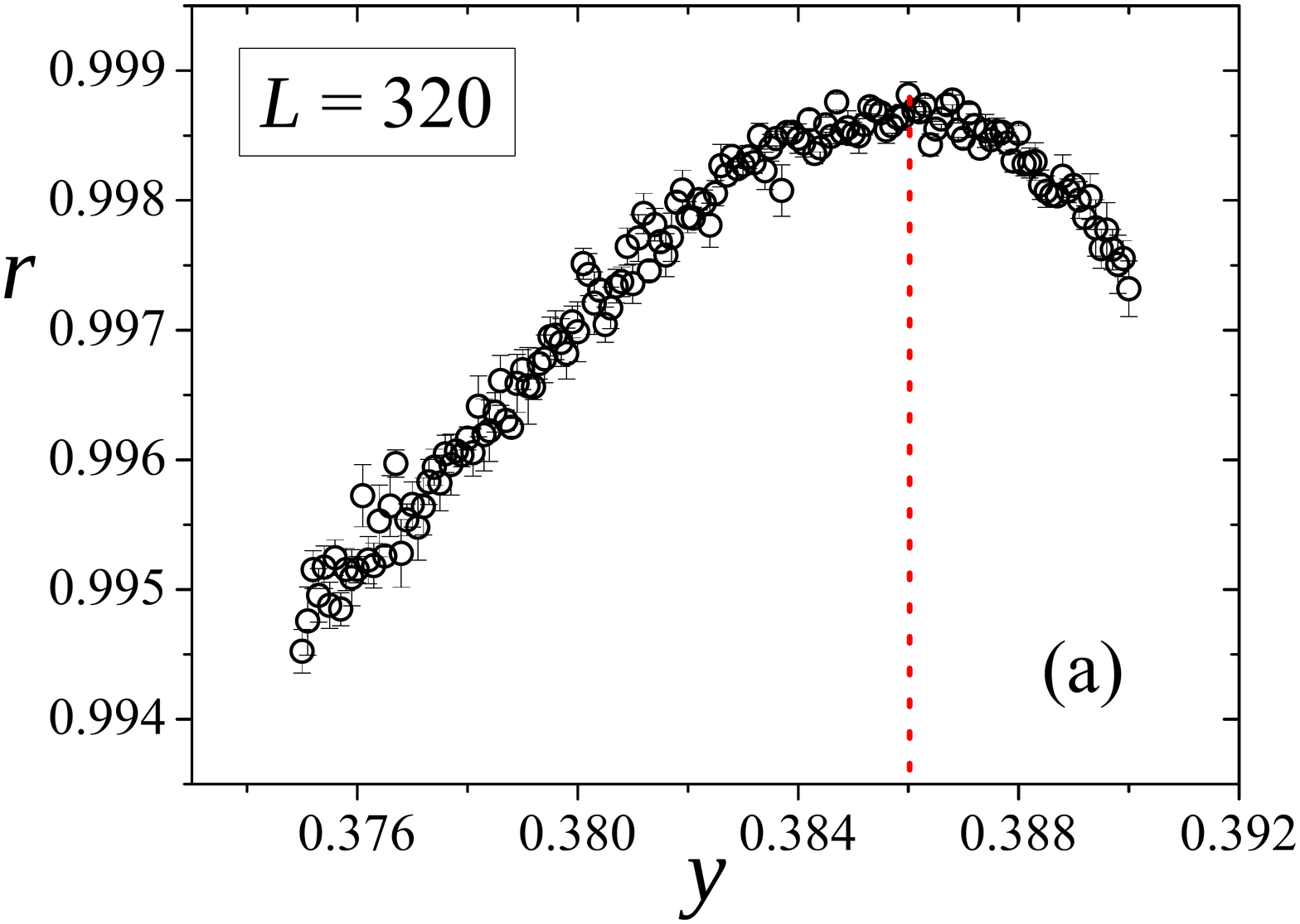} %
\includegraphics[width=\columnwidth]{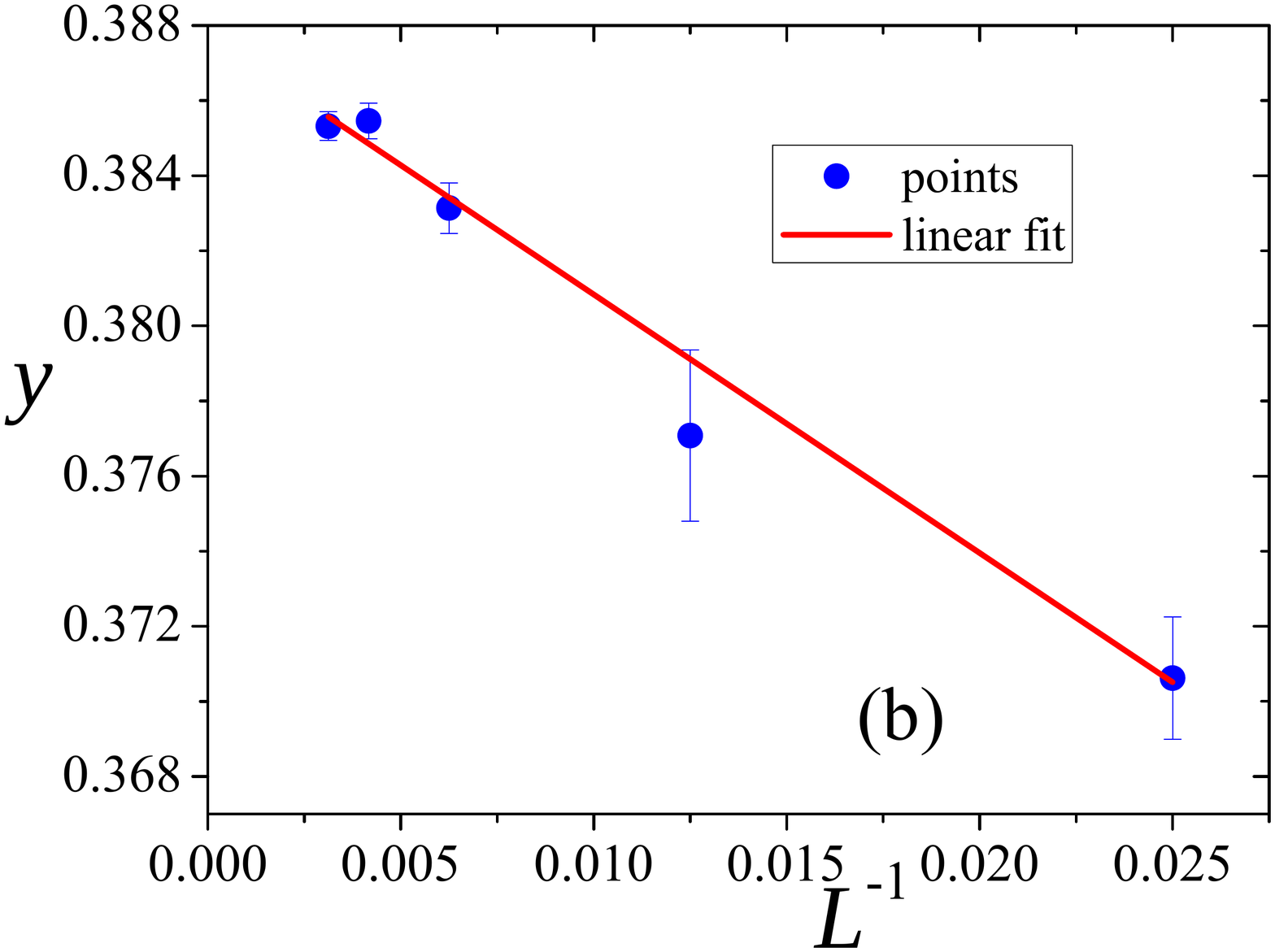}
\end{center}
\caption{Determination of second-order point using the density of $CO$
molecules through the curve $r\times y$. (a): Localization of the second
order point for $L=320$. (b) Extrapolation of $y\times 1/L$ for different
the lattice sizes used in this paper.}
\label{Fig:Second_order_cox}
\end{figure}

Firstly, we focused our attention to the candidates to the second-order
point. For each lattice size, we obtain the maximum for different seeds by
taking an average. Finally, an extrapolation was performed to take into
account effects of finite size. Figure \ref{Fig:Second_order_cox} shows the
localization of the second-order point $y_{1}$ for $\rho _{CO}$. The
behavior of $r$ versus $y$ is shown in Fig. \ref{Fig:Second_order_cox}(a)
and \ref{Fig:Second_order_cox}(b) presents the extrapolation $y$ versus $1/L$%
. 

\begin{figure}[th]
\begin{center}
\includegraphics[width=\columnwidth]{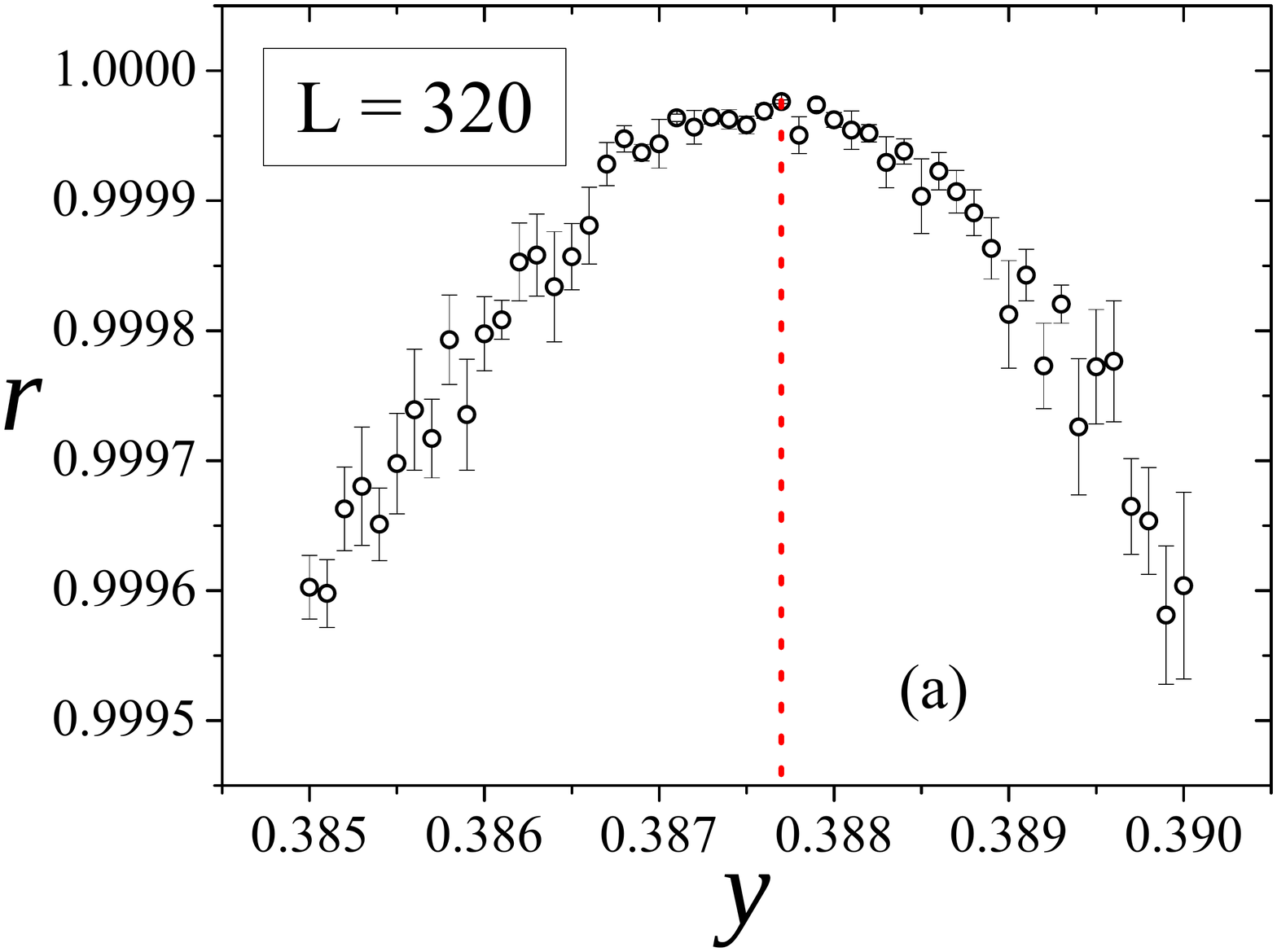} %
\includegraphics[width=\columnwidth]{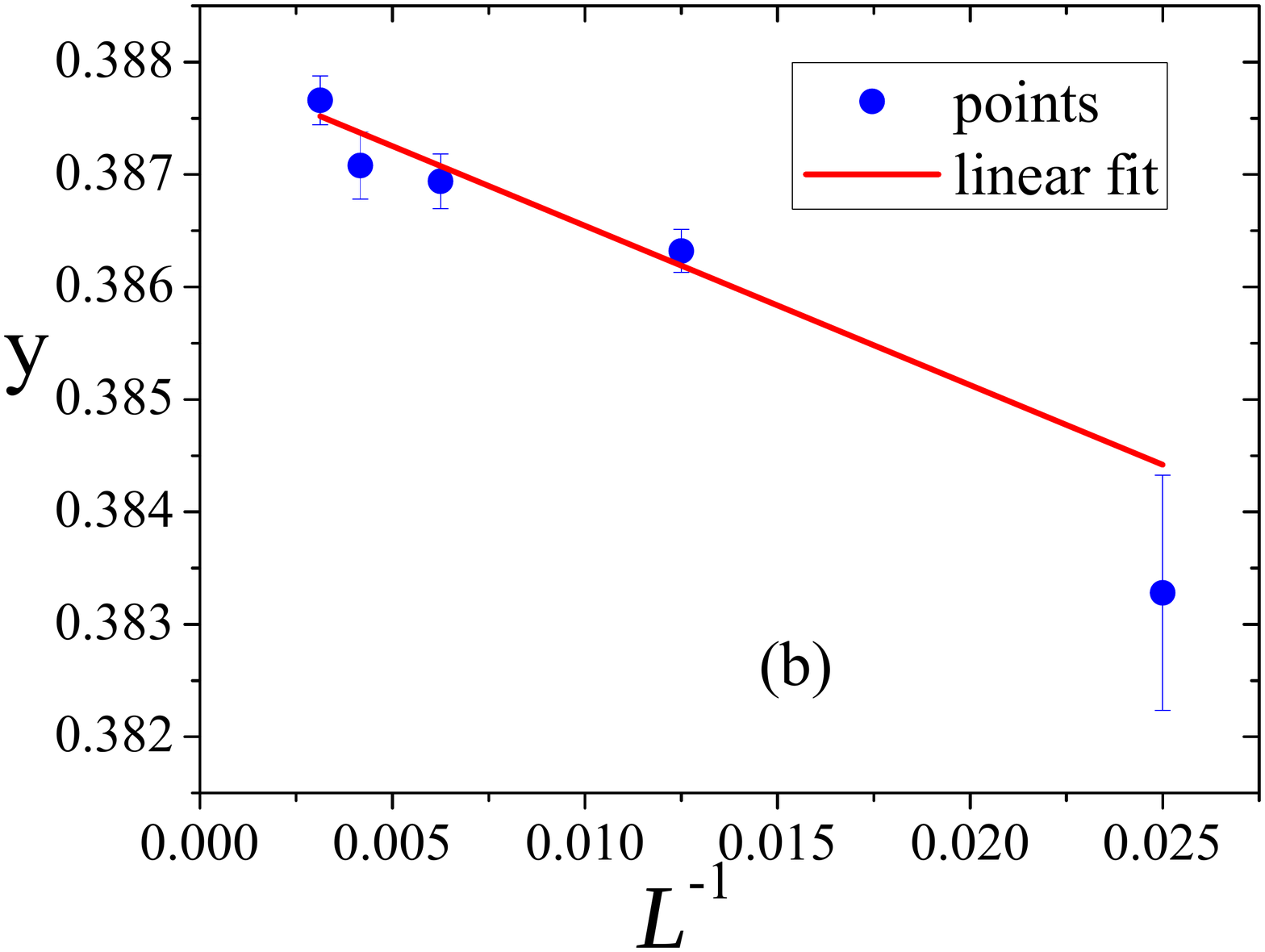}
\end{center}
\caption{Determination of second-order point using the density of empty
sites through the curve $r\times y$. (a): Localization of the second order
point for $L=320$. (b) Limit procedure used to determine $y_{1}$ when $%
1/L\rightarrow 0$.}
\label{Fig:Second_order_empty}
\end{figure}

We also study the coefficient of determination considering $\rho _{V}$ as
order parameter in order to check the efficiency of the method to obtain the
second-order point of the model. In Fig. \ref{Fig:Second_order_empty} we
present both behavior of $r$ versus $y$ for $L=320$ (plot (a)) and the
estimates for different lattice sizes along with the limit procedure (plot
(b)) using the density of vacant sites. 

The values for $y_{1}$ (second-order critical point) obtained in this paper
are presented in Table \ref{Table:critical_parameters}. We can observe an
excellent agreement with the first estimate obtained in literature.

\begin{table}[tbp]
\centering
\begin{tabular}{lll}
\hline
& Second-order point & Upper Spinodal point \\ \hline
$\rho_{CO}$ & $0.3877(5)$ & $0.52738(14)$ \\ 
$\rho_{V}$ & $0.3879(2)$ & $0.52764(12)$ \\ 
Literature & $\cong 0.3874\ $\cite{voigt1997} & $0.5270(5)$ \cite{Albano2001}
\\ \hline
\end{tabular}%
\caption{Our results for the second-order and spinodal points obtained by
the method of optimization of power laws. Our estimates are in excellent
agreement with literature}
\label{Table:critical_parameters}
\end{table}

\begin{figure}[ht!]
\begin{center}
\includegraphics[width=\columnwidth]{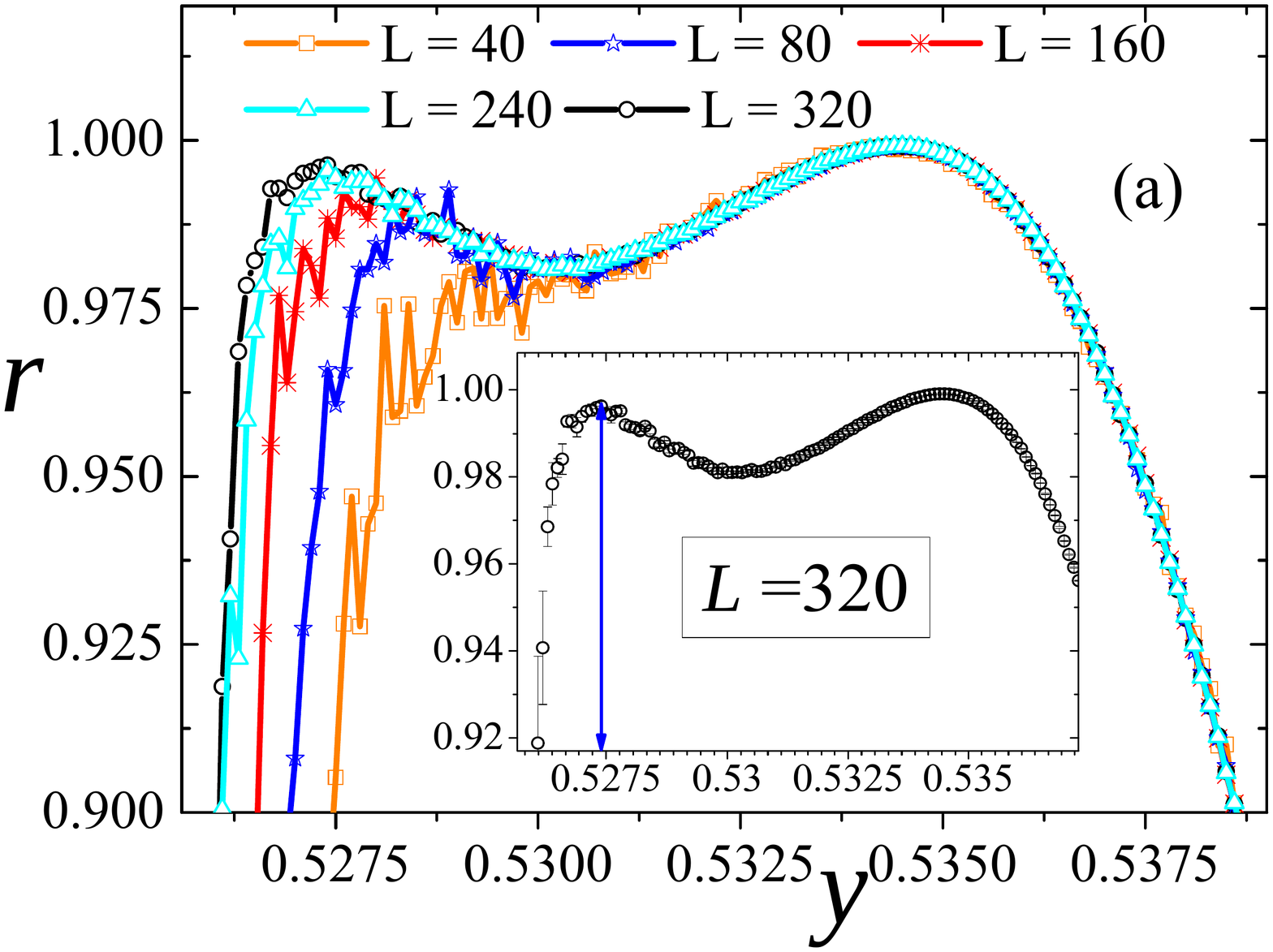} %
\includegraphics[width=\columnwidth]{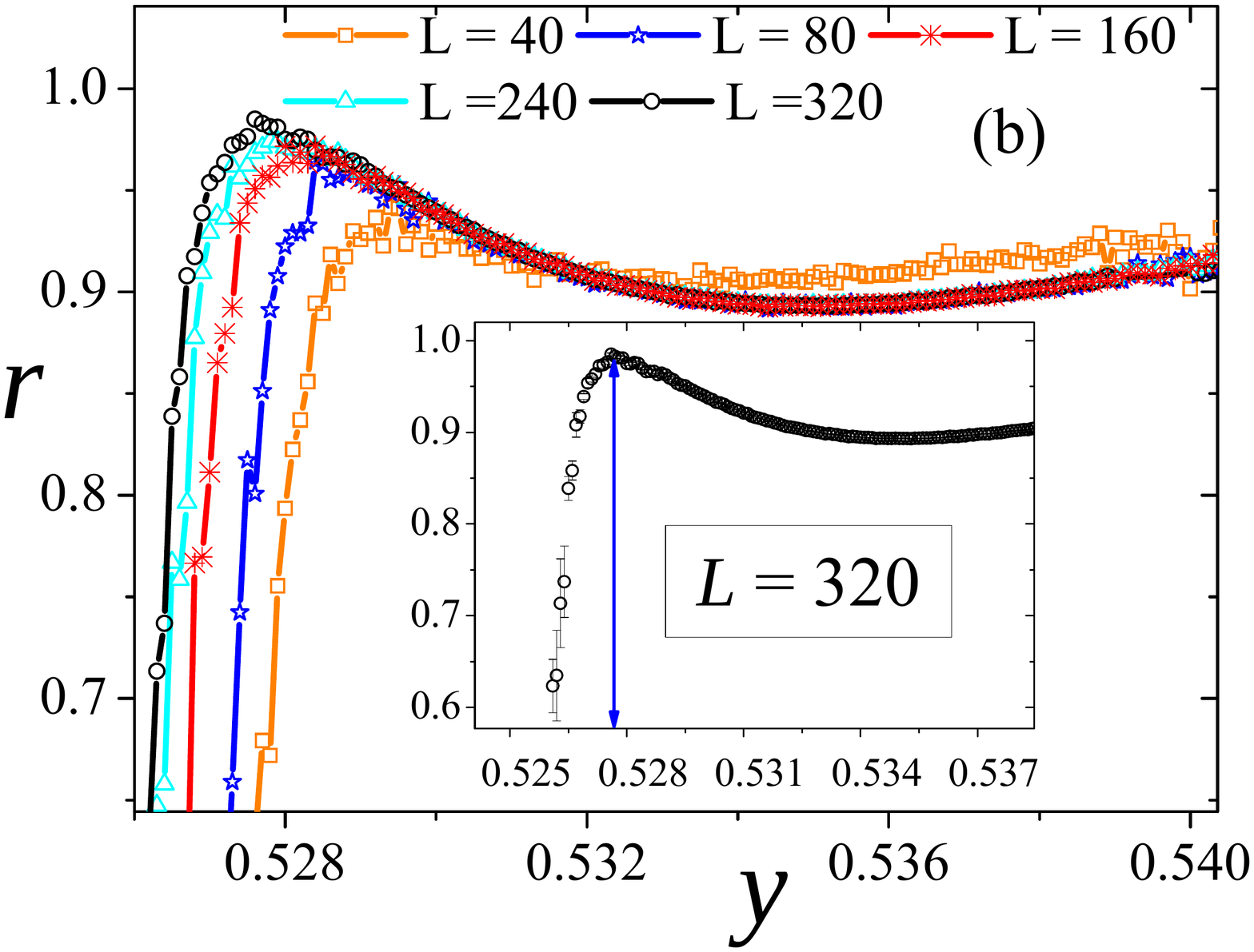}
\end{center}
\caption{Determination of the upper spinodal point. (a): Using the density
of $CO$ molecules (b) Using the density of empty sites. The plots show the
curves $r\times y$ for different lattice sizes. The inset one shows the plot
only with our larger lattice $L=320$ with error bars indicating the
pronounced (first) peak which corresponds to the upper spinodal point.}
\label{Fig:First_order}
\end{figure}

Now, we focus our attention to the upper spinodal point $y_2^{up}$
previously predicted by other authors (see \cite{Albano2001}). Plot (a) of
Fig. \ref{Fig:First_order} shows $r$ $\times \ y$ for the density of $CO$
molecules for an isolated region which is candidate to contain a weak
first-order transition. This figure presents all lattice sizes considered in
this paper. For clarity, the error bars are not shown in the main plots.

In Fig. \ref{Fig:First_order}(a) we can see two different `hills'. It is
important to observe that the second hill does not change for different
lattice sizes. Whereas we know that finite size scaling involving
first-order transitions or even their spinodal points (in the case of weak
first-order transitions) are notable, we particularly concentrate our
attention in the first point which moves to the left as $L$ increases. Since
we have large fluctuations of smaller sizes, we do not use an extrapolation
here and instead we directly calculate the $y$ that maximizes $r$ for five
different seeds in our largest lattice ($L=320$) which is shown in the inset
plot in Fig. \ref{Fig:First_order}(a) with the appropriated error bars.
Surprisingly we find $y_{2}^{up}=0.52738(14)$ which exactly matches what is
found in the literature for the upper spinodal point of the first-order
transition point \cite{Albano2001}. It is expected that spinodal points
(also called pseudo-critical points) behave as critical points as shown by
Schulke and Zheng \cite{Schulke2001} in the context of short-time dynamics.
This excellent agreement led us to investigate the density of empty sites,
as shown in plot (b) of Fig. \ref{Fig:First_order}, and our result, $%
y_{2}^{up}=0.52764(12)$, is in excellent agreement with our previous
estimate for the upper spinodal point. Our main results are resumed in Table %
\ref{Table:critical_parameters}. Here, it is important to mention that the
second hill does not appear and its no size dependence seems to be correctly
disregarded in the first situation.

In the next section we obtain the critical exponents of the second-order
critical point. To our knowledge, this is the first time that the considered
exponents are computed with short-time MC method.

\subsection{Results II: Critical exponents}

\label{subsec:resultsII}

Finally, we perform short-time MC simulations to obtain the dynamic and
static critical exponents of the ZGB model. In our simulations, we consider
square lattices of linear sizes $L=80$, $160$, $240$, and $320$ in order to
account for finite size effects, and the density of $CO$ molecules is
considered as the order parameter of the model.

Here, we considered $N_{MC}=500$ MC steps in the study of the time evolution
given by Eqs. (\ref{eq_pl1}) and (\ref{eq_pl2}) and $N_{MC}=1500$ MC steps
when considering the Eq. (\ref{eq_dt}). However, first 100 MC steps were
disregarded in the calculation of the exponents $\beta/\nu_{\parallel}$ and $%
\theta$. On the other hand, to obtain the exponent $1/\nu_{\parallel}$, we
disregarded the 500 MC steps at the beginning of the simulation. In
addition, to estimate these exponents with precision, we perform huge
simulations with $N_{run}=10000$ runs.

\begin{figure}[ht!]
\begin{center}
\includegraphics[width=1\columnwidth]{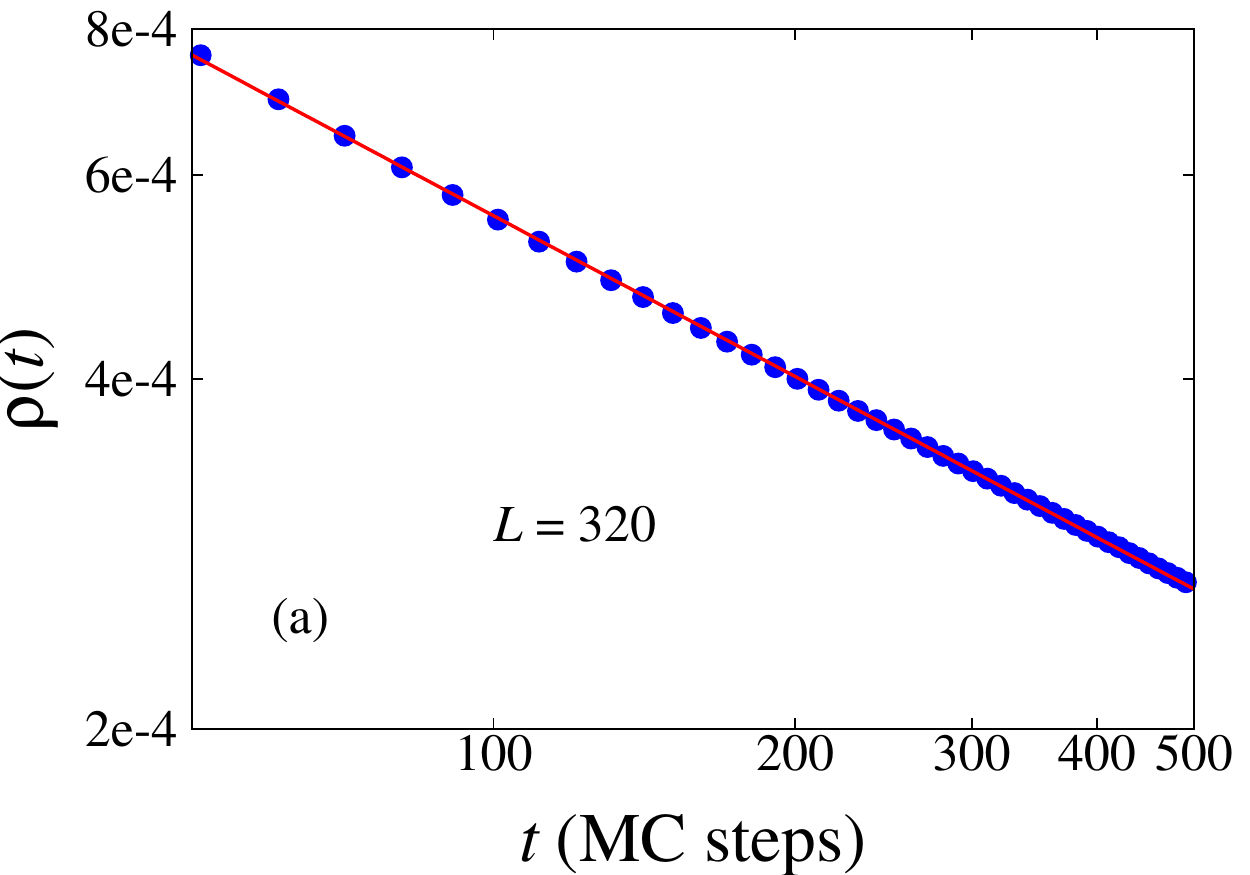} \includegraphics[width=1%
\columnwidth]{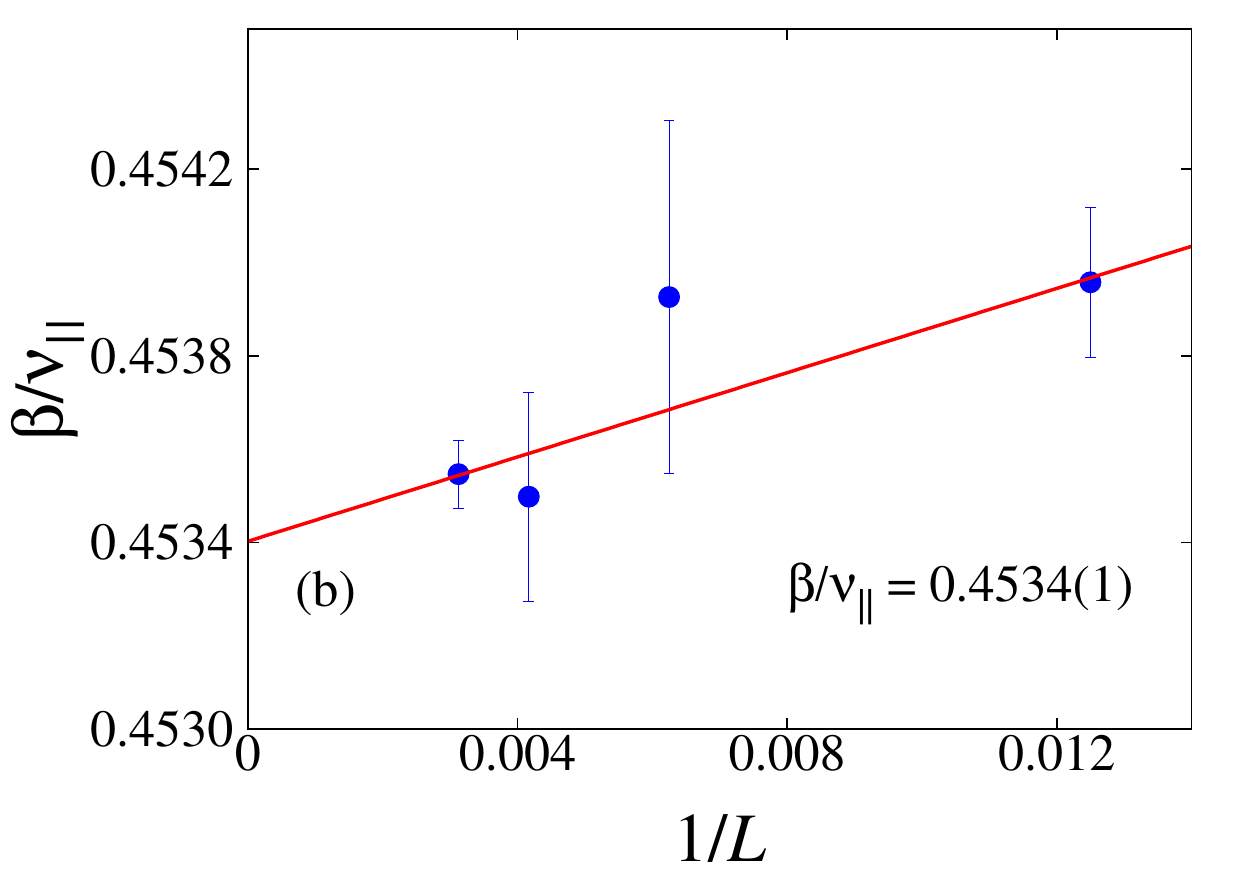}
\end{center}
\caption{(a) Time evolution of $\protect\rho(t)$ when the initial lattice is
completely empty. (b) Limit procedure $L\rightarrow \infty$ to obtain $%
\protect\beta/\protect\nu_\parallel$ in the thermodynamic limit.}
\label{fig_coxv}
\end{figure}

Figure \ref{fig_coxv}(a) shows de behavior of Eq. (\ref{eq_pl1}) in $\text{%
log}-\text{log}$ scale for $L=320$. The error bars are smaller then the
symbols. 

Through the linear fit of this curve we obtain $\beta /\nu _{\parallel
}=0.4535(1)$. In order to take into account the effects of finite size, we
also simulate the system with other lattice sizes. In Fig. \ref{fig_coxv}%
(b), we show the limit procedure $L\rightarrow \infty $ used to reach the
thermodynamic limit. In Table \ref{tab_coxv} we show our results along with
the estimates obtained for $L\rightarrow \infty $. {\ 
\begin{table}[h]
\begin{tabular}{cccccc}
\hline
Exponent & $L=80$ & $L=160$ & $L=240$ & $L=320$ & $L\rightarrow \infty $ \\ 
\hline
$\beta /\nu _{\parallel }$ & 0.4539(2) & 0.4539(4) & 0.4535(3) & 0.4535(1) & 
0.4534(1) \\ \hline
\end{tabular}%
\caption{Static critical exponent $\protect\beta /\protect\nu _{\parallel }$
for different lattice sizes as well as the extrapolated value when $%
L\rightarrow \infty $.}
\label{tab_coxv}
\end{table}
}

\begin{figure}[ht!]
\begin{center}
\includegraphics[width=1\columnwidth]{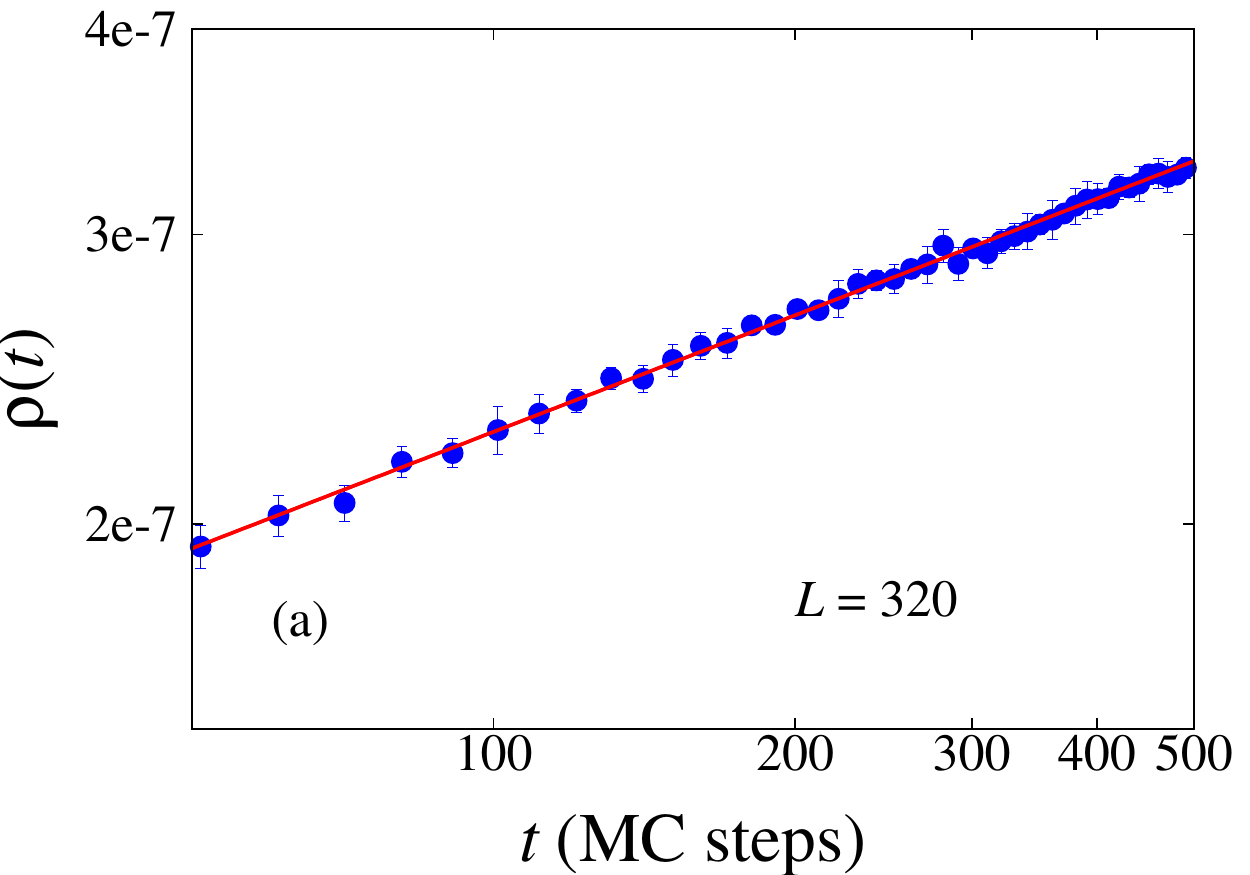} \includegraphics[width=1%
\columnwidth]{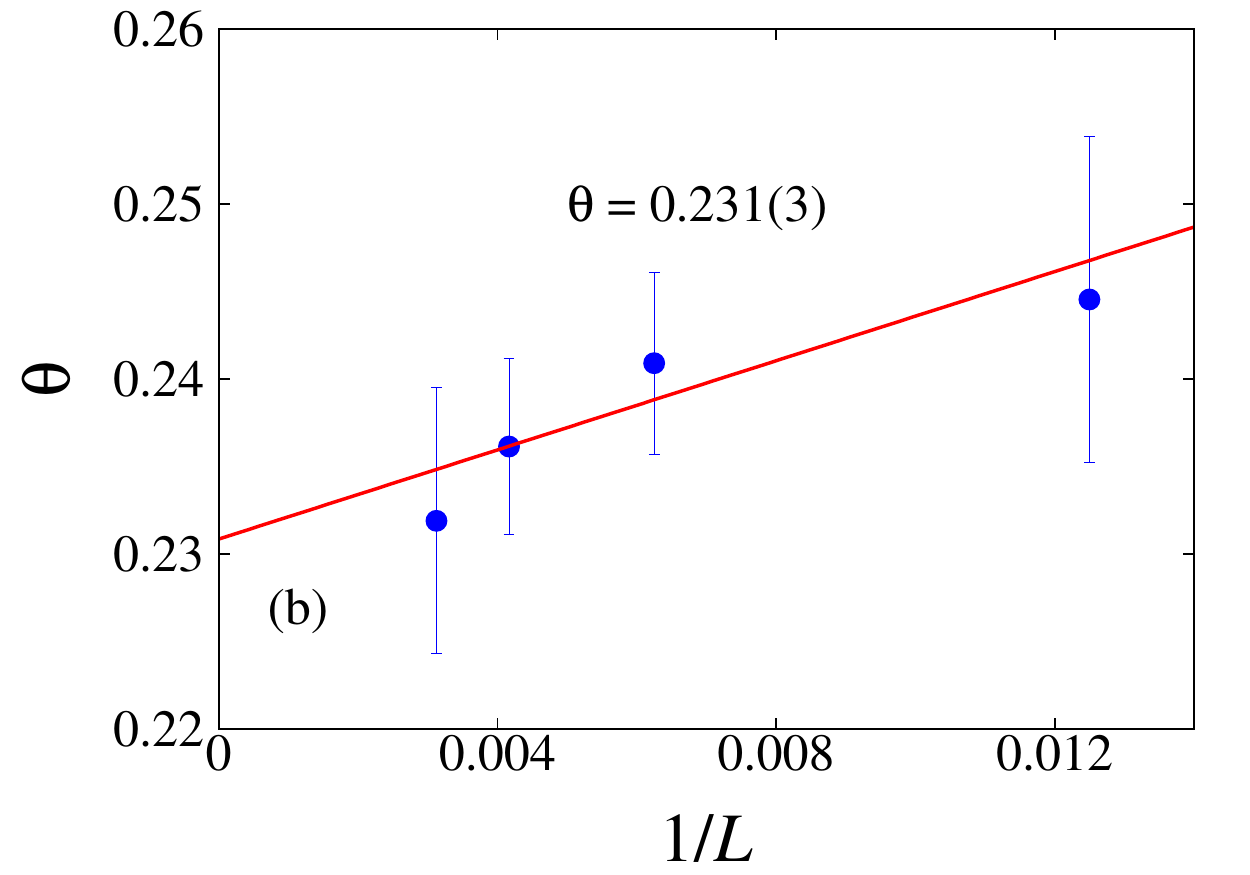}
\end{center}
\caption{(a) Time evolution of $\protect\rho(t)$ when the initial lattice is
completely filled with $O$ atom but a unique random site which remains
empty. (b) Limit procedure $L\rightarrow \infty$ to obtain the dynamic
exponent $\protect\theta$ in the thermodynamic limit.}
\label{fig_coxo}
\end{figure}

Figure \ref{fig_coxo}(a) shows de behavior of Eq. (\ref{eq_pl2}) in $\text{%
log}-\text{log}$ scale for $L=320$. The slope of this curve is the dynamic
critical exponent $\theta$. 
In Fig. \ref{fig_coxo}(b), we present our estimates for different lattice
sizes along with the limit procedure whose result when $L \rightarrow \infty$
is $\theta=0.231(3)$.

Table \ref{tab_coxo} summarizes our estimates for all considered lattices
(as presented in Fig. \ref{fig_coxo}(b)) and for $L\rightarrow \infty $. {\ 
\begin{table}[h]
\begin{tabular}{cccccc}
\hline
Exponent & $L=80$ & $L=160$ & $L=240$ & $L=320$ & $L\rightarrow \infty $ \\ 
\hline
$\theta $ & 0.245(9) & 0.241(5) & 0.236(5) & 0.232(8) & 0.231(3) \\ \hline
\end{tabular}%
\caption{Dynamic critical exponent $\protect\theta $ for different lattice
sizes as well as the extrapolated value when $L\rightarrow \infty $.}
\label{tab_coxo}
\end{table}
}

\begin{figure}[ht!]
\begin{center}
\includegraphics[width=1\columnwidth]{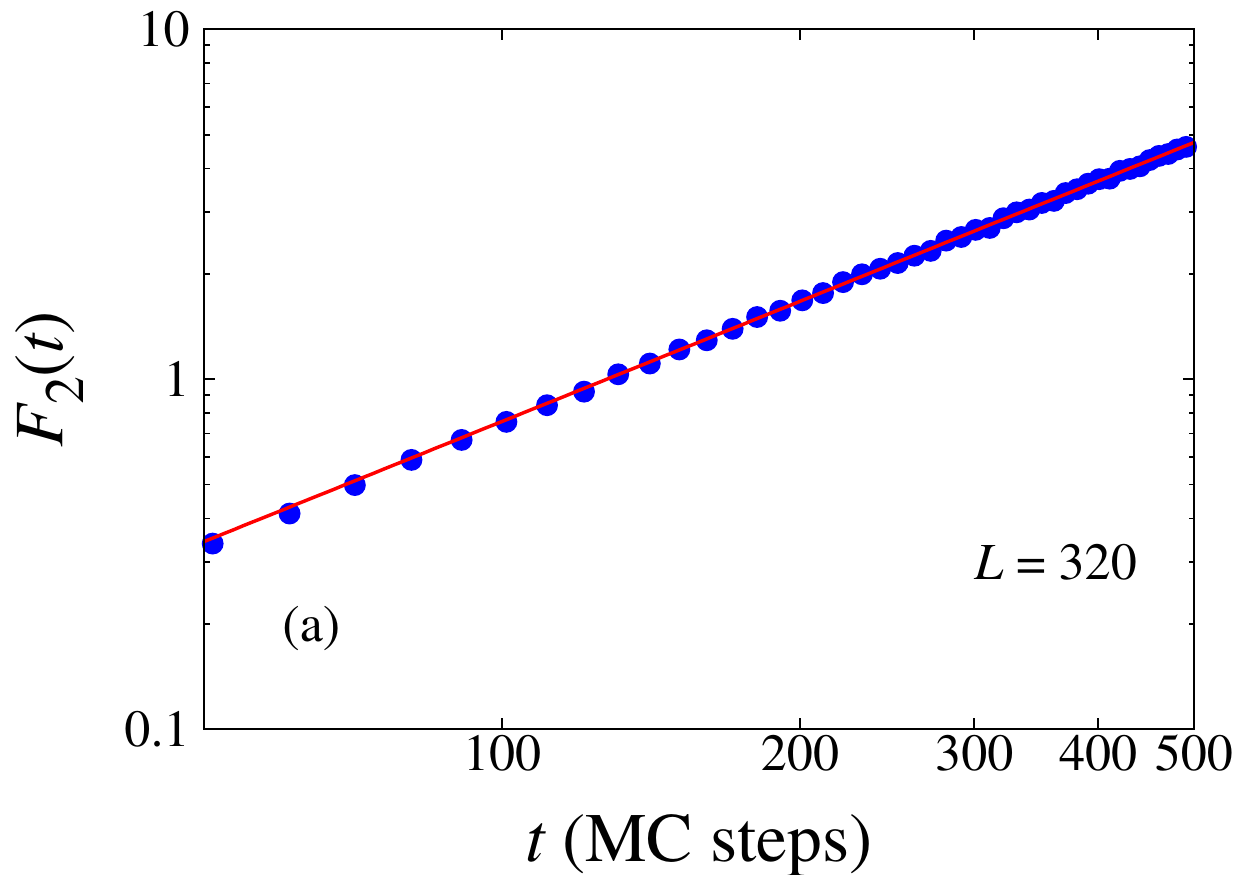} \includegraphics[width=1%
\columnwidth]{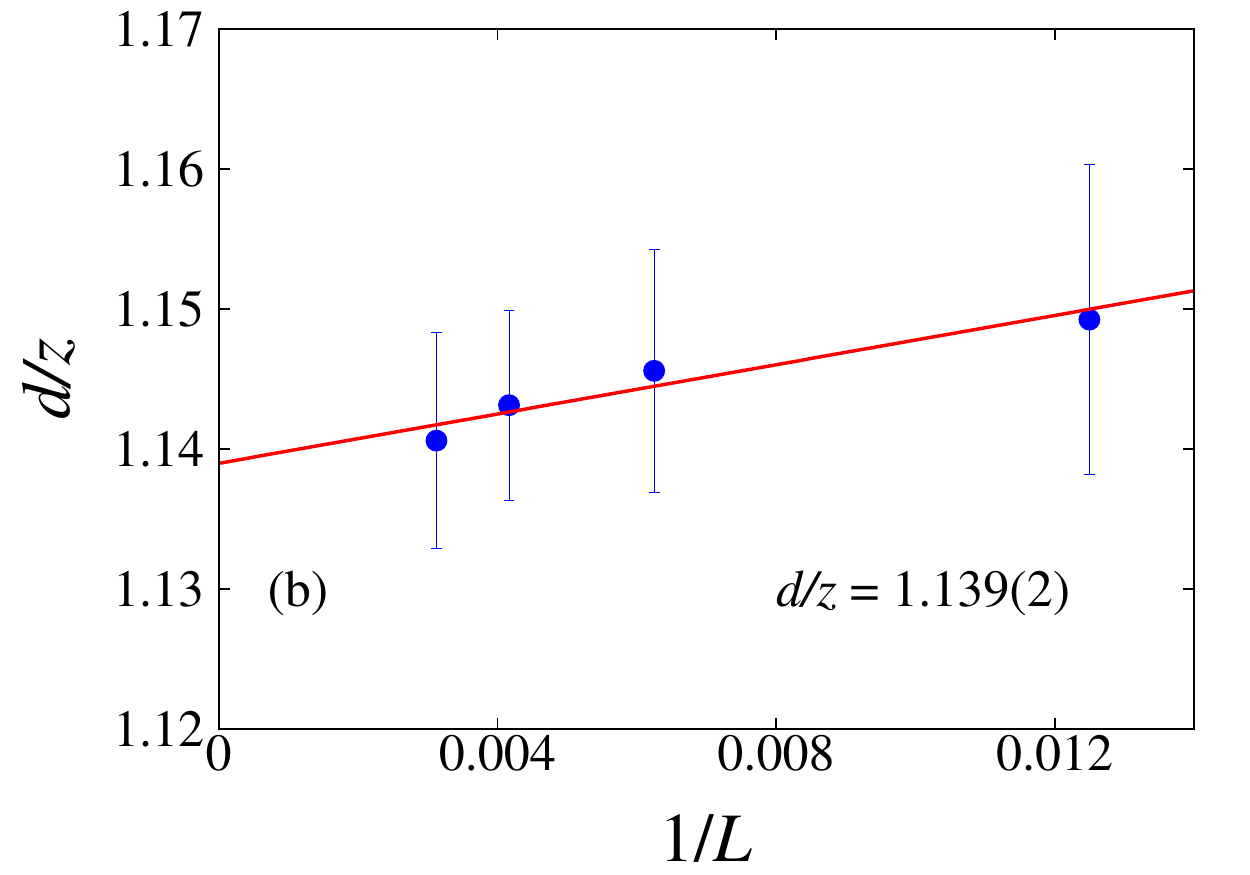}
\end{center}
\caption{(a) Time evolution of $F_2(t)$ in $\text{log}-\text{log}$ scale for 
$L=320$. The error bars are smaller than the symbols. (b) Limit procedure $%
L\rightarrow \infty$ to obtain the dynamic exponent $d/z$ in the
thermodynamic limit.}
\label{fig_f2}
\end{figure}

As mentioned above, the dynamic critical exponent $z$ can be obtained,
independently from other exponents, by considering the function $F_{2}(t)$
[Eq. (\ref{eq_f2})]. Figure \ref{fig_f2}(a) shows the time evolution in $%
\text{log}-\text{log}$ scale of $F_{2}(t)$ for the model when $L=320$. 
By following the same procedure as before, one can obtain the extrapolated
value (when $L \rightarrow \infty$) of the exponent $d/z$ through the limit
procedure shown in Fig. \ref{fig_f2}(b). Our estimates for the considered
lattice sizes as well as when $L \rightarrow \infty$ are presented in Table %
\ref{tab_f2}.

{\ 
\begin{table}[h]
\begin{tabular}{cccccc}
\hline
Exponent & $L=80$ & $L=160$ & $L=240$ & $L=320$ & $L\rightarrow \infty $ \\ 
\hline
$d/z$ & 1.149(11) & 1.146(9) & 1.143(7) & 1.141(8) & 1.139(2) \\ \hline
\end{tabular}%
\caption{Critical exponent $d/z$ for different lattice sizes as well as the
extrapolated value when $L\rightarrow \infty $.}
\label{tab_f2}
\end{table}
}

So far, we have already obtained the exponents $\beta/\nu_{\parallel}$, $%
\theta$, and $d/z$, where $z=\nu_{\parallel}/\nu_{\perp}$. If we are able to
estimate the exponent $\nu_{\parallel}$ independently, we can obtain all the
considered exponents separately. In order to obtain this exponent, we follow
the time evolution of the Eq. (\ref{eq_dt}) for different lattice sizes and
the final value is also obtained trough the extrapolation $%
1/\nu_{\parallel}\times 1/L$.

\begin{figure}[th]
\begin{center}
\includegraphics[width=1\columnwidth]{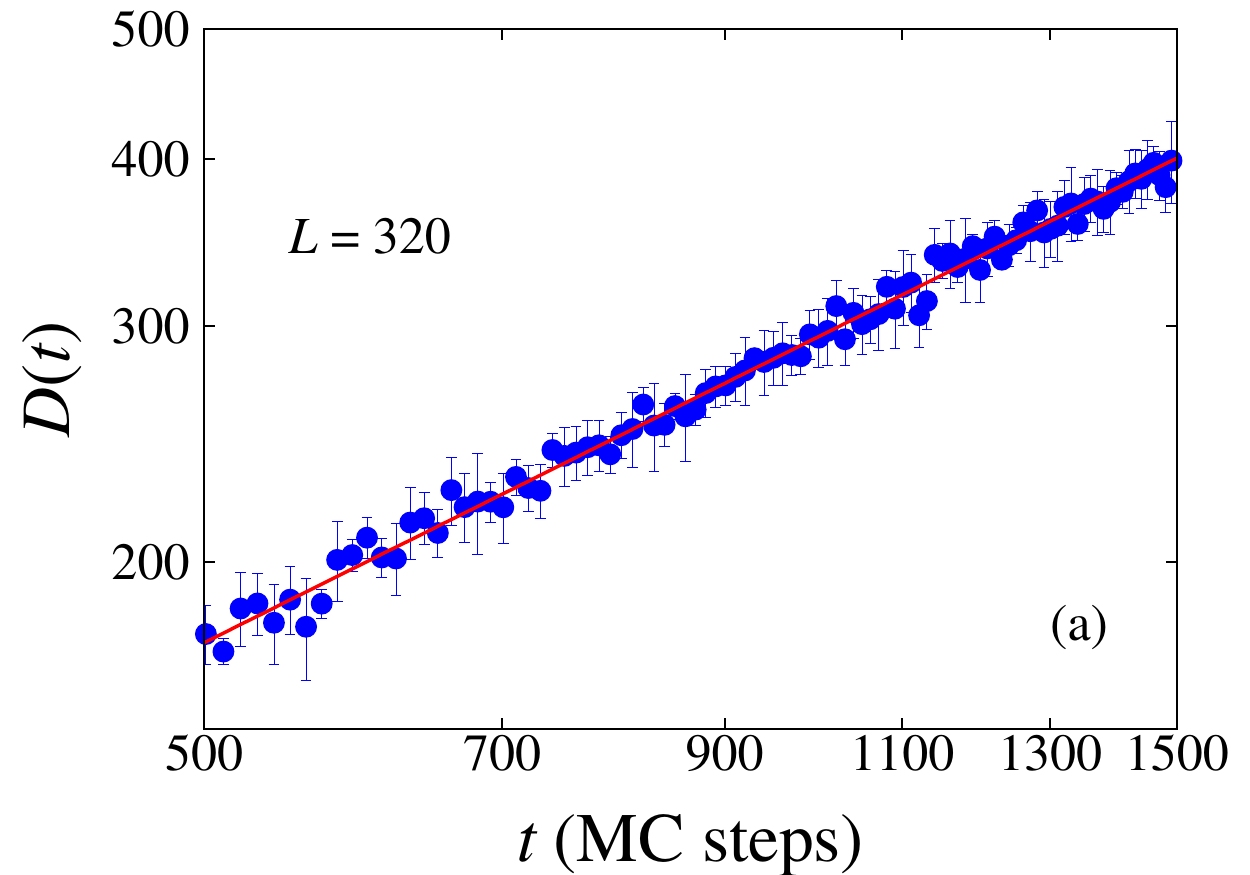} \includegraphics[width=1%
\columnwidth]{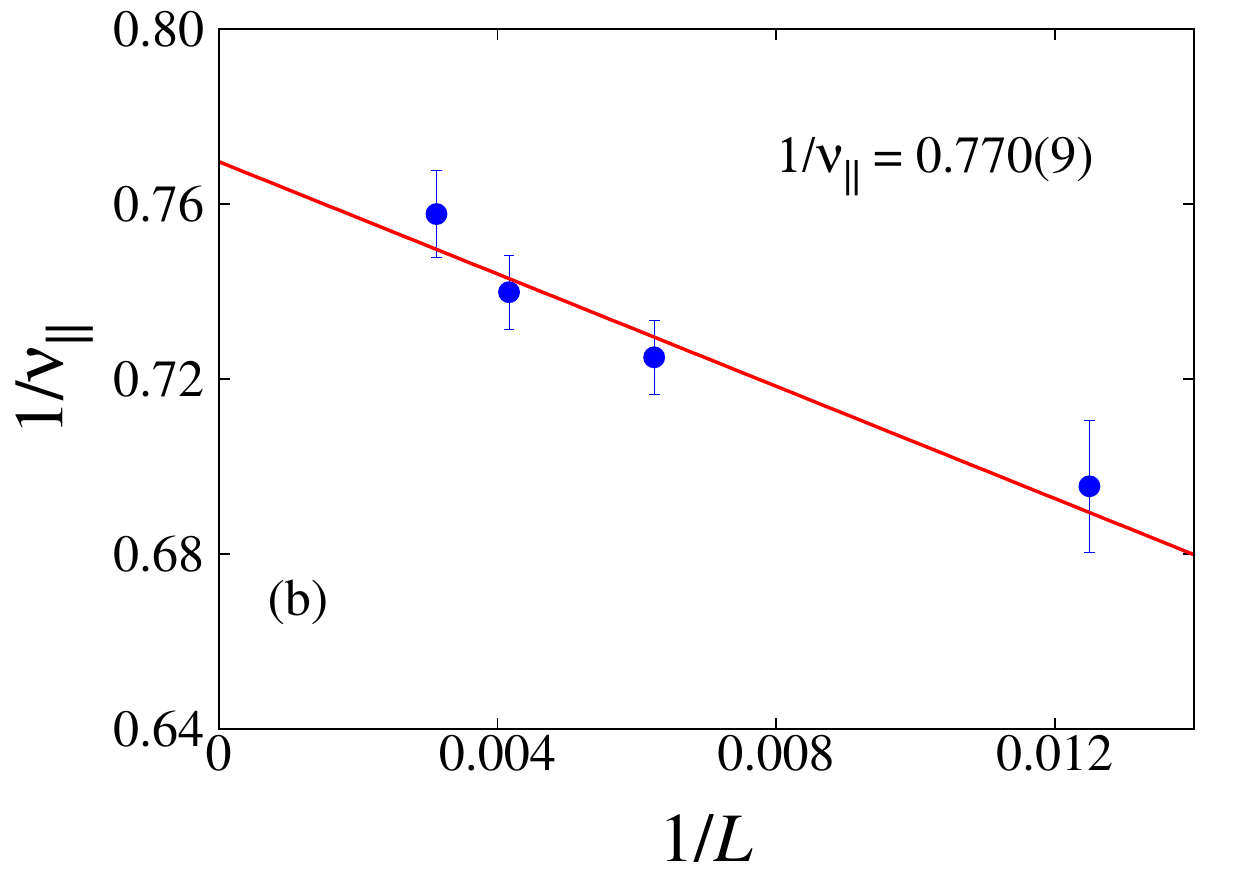}
\end{center}
\caption{(a) Time evolution of $D(t)$ in $\text{log}-\text{log}$ scale for $%
L=320$. (b) Limit procedure $L\rightarrow \infty $ to obtain the static
exponent $1/\protect\nu _{\parallel }$ in the thermodynamic limit.}
\label{fig_dvd}
\end{figure}

In Fig. \ref{fig_dvd}(a) we show the time evolution of $D(t)$ in $\text{log}-%
\text{log}$ scale for $L=320$ and the extrapolation is presented in Fig. \ref%
{fig_dvd}(b). 
Table \ref{tab_dvd} presents the exponent $1/\nu _{\parallel }$ obtained in
our simulations for different lattice sizes as well as its extrapolated
value.{\ 
\begin{table}[h]
\begin{tabular}{cccccc}
\hline
Exponent & $L=80$ & $L=160$ & $L=240$ & $L=320$ & $L\rightarrow \infty $ \\ 
\hline
$1/\nu _{\parallel }$ & 0.696(15) & 0.725(8) & 0.740(8) & 0.758(10) & 
0.770(9) \\ \hline
\end{tabular}%
\caption{Critical exponent $1/\protect\nu _{\parallel }$ for different
lattice sizes as well as the extrapolated value when $L\rightarrow \infty $.}
\label{tab_dvd}
\end{table}
}

Finally, with this set of critical exponents in hand, we are able to
estimate the static and dynamic critical exponents of the ZGB model
independently. Our results, presented in Table \ref{tab_final}, are in
complete agreement with estimates obtained previously for the model.

{\ 
\begin{table}[h]
\begin{tabular}{ccc}
\hline
Exponent & Our results & Other results \cite{voigt1997} \\ \hline
$\beta $ & 0.586(7) & 0.584(4) \\ 
$\nu _{\parallel }$ & 1.292(15) & 1.295(6) \\ 
$\nu _{\perp }$ & 0.736(10) & 0.734(4) \\ 
$\theta $ & 0.231(3) & 0.2295(10) \\ 
$z$ & 1.756(3) & 1.76(3) \\ \hline
\end{tabular}%
\caption{Static and dynamic critical exponents of the ZGB model}
\label{tab_final}
\end{table}
}

These results, along with the localization of the second-order phase
transition and the upper spinodal point of the ZGB model, show the
efficiency and reliability of short-time Monte Carlo simulations and the
coefficient of determination method in the study of systems without a
defined Hamiltonian and that possess absorbing states.

\section{Conclusions}

\label{conclusions}

In this work, we studied the phase transitions of the Ziff-Gulari-Barshad
(ZGB) by using an alternative method that optimizes the coefficient of
determination to localize the critical parameter of the second-order point
and an estimate of the upper spinodal point (one of pseudo critical points)
of the the weak first-order transition point of this model. To obtain these
points, we considered the density of $CO$ molecules ($\rho _{CO}$) and the
density of vacant sites ($\rho _{V}$) as order parameters of the model. In
this study, we found a second peak, on the right side of the upper spinodal
point that does not present effects of finite size and therefore was not
considered in this work. However, this point could be subject of further
investigation in order to clarify its meaning and relationship with the
first-order phase transition of he model. Moreover, we also obtain the
critical exponents of the second-order point by using time-dependent
simulations. The exponents $\beta $, $\nu _{\parallel }$, $\nu _{\perp }$, $z
$, and $\theta $ were obtained independently from the power laws. Our
results are in excellent agreement with previous results. The methodology
developed in this paper can be easily applied to the other surface reaction
models by including desorption, impurities or even mobility of molecules.

\section*{Acknowledgments}   
This research work was in part supported financially by CNPq (National
Council for Scientific and Technological Development)

\end{document}